\documentclass[aps,prl,showkeys,showpacs,amssymb,cite,
amsfonts,epsf,preprintnumbers,nofootinbib,superscriptaddress]{revtex4}

\usepackage[dvips]{graphicx}
\usepackage{bm,latexsym,amsmath,amssymb,amsfonts,color}

\newcommand{\be}{\begin{equation}}
\newcommand{\ee}{\end{equation}}
\newcommand{\bear}{\begin{eqnarray}}
\newcommand{\eear}{\end{eqnarray}}
\newcommand{\ba}{\begin{array}}
\newcommand{\ea}{\end{array}}
\newcommand{\nn}{\nonumber}

\begin{document}

\title{Precursor of Inflation}

\author{Inyong Cho}
\email{iycho@seoultech.ac.kr}
\affiliation{Institute of Convergence Fundamental Studies \& School of Liberal Arts,
Seoul National University of Science and Technology, Seoul 139-743, Korea}
\author{Hyeong-Chan Kim}
\email{hckim@ut.ac.kr}
\affiliation{School of Liberal Arts and Sciences, Korea National University of Transportation, Chungju 380-702, Korea}
\author{Taeyoon Moon}
\email{tymoon@sogang.ac.kr}
\affiliation{Institute of Basic Sciences and Department of Computer Simulation, Inje University, Gimhae 621-749, Korea}

\begin{abstract}
We investigate a nonsingular initial state of the Universe
which leads to inflation naturally.
The model is described by a scalar field with a quadratic potential
in Eddington-inspired Born-Infeld gravity.
The curvature of this initial state is given by the mass scale of the scalar field
which is much smaller than the Planck scale.
Therefore, in this model, quantum gravity is not necessary
in understanding this pre-inflationary stage,
no matter how large the energy density becomes.
The initial state in this model evolves eventually
to a long inflationary period which is similar to the usual chaotic inflation.
\end{abstract}
\maketitle

``How did the Universe begin?"
This is one of the oldest questions of the human being.
After the discovery of the Hubble's law in 1929,
people started to believe that the Universe originated from big bang.
However, the big-bang scenario soon confronted with obstacles
such as the flatness, horizon, and monopole problems.
The inflationary  scenario was introduced in 1980
to resolve these problems.
One of the simplest model of the inflation scenario must be the chaotic inflation
described by a scalar field with a quadratic potential,
\be \label{S:chaotic}
S_M = \int d^4 x \sqrt{-|g_{\mu\nu}|}
\left[ -\frac12 g_{\mu\nu} \partial^\mu\phi \partial^\nu \phi -V(\phi) \right],
\qquad
V(\phi) = \frac{m^2}{2} \phi^2,
\ee
where $|g_{\mu\nu}|$ represents the determinant of $g_{\mu\nu}$.
The Universe undergoes inflation during the slow-roll evolution of the scalar field,
during which the scalar field and the scale factor are approximated by
\be\label{slowsol}
\phi(t) \approx \phi_i +\sqrt{2/3}m t,
\qquad
a(t) \approx a_i\; e^{ \frac{1}{4} [\phi_i^2-\phi^2(t)]},
\ee
where $\phi_i <0$ is the value of the inflaton field in the beginning of inflation.
(We assume that  $\phi$ rolls down the potential $V(\phi)$
in the region of $\phi<0$, and set $8\pi G=1$ in this letter.)
This slow-roll solution is known to be an attractor.
For $N\sim 70$ $e$-foldings, $|\phi_i| \gtrsim 10$ is required,
and $m \sim 10^{-5}$ from observational data.

Even with the success of the inflationary scenario,
it connotes problems
such as the fine-tuning~\cite{Boyle:2005ug}
and the low-entropy initial state~\cite{Page:1983uh}.
In addition, we still do not know how inflation appears
and what happened before and at the early stage of inflation.
Due to the high-curvature scale in the theory of general relativity (GR),
these issues inevitably require the introduction of quantum gravity.

In this letter, we show that the Eddington-inspired Born-Infeld (EiBI) theory of gravity,
recently proposed by Ban\~{a}dos and Ferreira~\cite{Banados:2010ix},
provides a {\it natural origin of inflation} without introducing quantum gravity.
For the same matter field in Eq.~\eqref{S:chaotic},
at the high-energy scale in EiBI theory,  the curvature scale remains finite
while the Universe undergoes a pre-inflationary exponential expansion,
followed by the ordinary chaotic inflationary period.

The EiBI theory of gravity is described by the action,
\begin{eqnarray}\label{maction}
S_{{\rm EiBI}}=\frac{1}{\kappa}\int
d^4x\Big[~\sqrt{-|g_{\mu\nu}+\kappa
R_{\mu\nu}(\Gamma)|}-\lambda\sqrt{-|g_{\mu\nu}|}~\Big]+S_M(g,\Phi),
\end{eqnarray}
where $\lambda$  is a dimensionless parameter related with the
cosmological constant by $\Lambda = (\lambda -1)/\kappa$, and
$\kappa$ is the only additional parameter of the theory. In this
theory the metric $g_{\mu\nu}$ and the connection
$\Gamma_{\mu\nu}^{\rho}$ are treated as independent fields (Palatini
formalism). The Ricci tensor $R_{\mu\nu}(\Gamma)$ is evaluated
solely by the connection, and the matter field $\Phi$ is coupled
only to the gravitational field $g_{\mu\nu}$. The merit of EiBI
theory is that it is equivalent to GR in the vacuum (or with
cosmological constant). Therefore, the results of astronomical
experiments such as the light deflection by a star are still viable.
In addition, it predicts a singularity-free initial state of the
Universe filled with perfect fluid~\cite{Banados:2010ix,Cho:2012vg}.
Subsequent work has been investigated in the cosmological and
astrophysical aspects in
Refs.~\cite{Pani:2011mg,Pani:2012qb,DeFelice:2012hq,Avelino:2012ge,Avelino:2012qe,Casanellas:2011kf,
EscamillaRivera:2012vz,Avelino:2012ue,Liu:2012rc,Delsate:2012ky,Pani:2012qd,Cho:2013usa}.

Let us consider the Universe driven the scalar field described by the action \eqref{S:chaotic} in
the spatially flat homogeneous and isotropic spacetime,
\begin{eqnarray}
g_{\mu\nu} dx^\mu dx^\nu = -dt^2 + a^2(t) d{\bf x}^2.
\end{eqnarray}
The scalar-field equation is given by
\begin{equation} \label{eom:phi}
\ddot \phi +3 H \dot \phi + V'(\phi) =0,
\end{equation}
where the prime denotes the differentiation with respect to $\phi$.
The Hubble parameter \cite{Cho:2013usa,Scargill:2012kg} is given by
\begin{align}\label{H}
H \equiv & \frac{\dot a}{a} =
    \frac{1}{\left(\lambdabar + V \right)^2 + \dot\phi^4/2}
\left\{ -\frac{1}{2}\left(\lambdabar+V + \frac{\dot \phi^2}{2}\right)
     V'(\phi)\dot \phi  \pm\frac{1}{\sqrt{3}}\left(\lambdabar+V-\frac{\dot \phi^2}2\right) \times  \right.  \nn\\
     & \left. \left[\left(\lambdabar+V +\frac{\dot \phi^2}{2}\right)^{3/2}
\left(\lambdabar+V-\frac{\dot \phi^2}2\right)^{3/2}
-\frac{1}{\kappa}\left(\lambdabar+V + \frac{\dot \phi^2}{2}\right)
    \left( \lambdabar+V -\dot \phi^2 \right)
     \right]^{1/2}  \right\},
\end{align}
where $\lambdabar \equiv\lambda/\kappa$.
(We shall consider the case of $\kappa, \lambda >0$ in this work.)
A distinct feature of this equation is that the curvature scale
is not simply proportional to the energy scale.
One of the consequences of this feature is the existence of
a nongravitating-dynamical field solution investigated in Ref.~\cite{Cho:2013usa}.
The other is what we study in this work,
the non-quantum birth of the inflationary Universe without beginning.

If we consider the evolution of the scalar field with the first slow-roll condition,
$\dot\phi^2 \ll \lambdabar+V(\phi)$,
the Hubble parameter is proportional to the scalar field,
$H \approx \sqrt{[V(\phi)+\Lambda]/3}= (m/\sqrt{6})|\phi|$,
which is the same as in GR (we assume $\Lambda=0$).
Then the scalar-field equation without ignoring $\ddot\phi$ term becomes
\be\label{ddf}
\ddot \phi +3 H \dot \phi +m^2 \phi =0
\quad\Rightarrow\quad
\frac{\frac12\frac{d \dot\phi^2}{dt}}{|\dot \phi|-\sqrt{2/3}m} \approx \sqrt{\frac32} \frac{m}{2} \frac{d \phi^2}{dt}.
\ee
This equation can be integrated to give
\be\label{df}
|\dot\phi| + \sqrt{2/3}m \log\big| |\dot\phi|-\sqrt{2/3}m\big|
\approx \sqrt{3/8} m (\phi^2 -\phi_0^2),
\ee
where, $\phi_0 \sim \phi_i$.
If $|\phi| < |\phi_0|$ [$\ddot\phi$ term in Eq.~\eqref{ddf},
or the first term in Eq.~\eqref{df} is subdominant; this is the second slow-roll condition],
$|\dot\phi| \approx \sqrt{2/3}m$ and
the Universe evolves along the attractor trajectory.
If $|\phi| > |\phi_0|$ (only the first slow-roll condition is satisfied),
the first term in Eq.~\eqref{df} becomes
dominant to the second term,
so we have $\dot\phi \propto \phi^2$.
Therefore, the $\dot\phi$ increases rapidly as we go back in time,
while $\phi$ climbs up the potential.
The energy scale also increases rapidly due to this dynamical behavior,
and arrives at the Planck scale soon.
In GR, the quantum gravity is required beyond this point.
However, we shall show below that it is not the story in the EiBI theory,
in which there exists an upper limit on the value of the field velocity $\dot\phi$.
In addition, the curvature scale is not simply proportional to the energy scale
as we can see from Eq.~\eqref{H}.

The upper limit of $\dot\phi$ comes from the requirement
that $H$ be real-valued.
This means from Eq.~\eqref{H} that
the value of $\lambdabar -\dot\phi^2/2 +V(\phi) \equiv \lambdabar - p$ should be non-negative.
The maximum value of $\dot\phi^2$ is achieved when
\begin{equation}\label{cond}
\frac12 \dot \phi^2 - V(\phi) -\lambdabar = 0 ,
\end{equation}
which we shall call the {\it maximal pressure condition} (MPC).
For the perfect fluid investigated in Refs.~\cite{Banados:2010ix,Cho:2012vg},
MPC is responsible for the {\it nonsingular initial state} of the Universe.
Equation \eqref{cond} can be rewritten as\footnote{We shall consider
only the positive signature since
$\dot\phi=-U$ can be obtained by
making the replacement $t\to-t$ in Eq.~\eqref{dphi}.}
\begin{equation}\label{dphi}
\dot\phi =  U(\phi)
\quad\Rightarrow\quad
\ddot\phi = U(\phi)U'(\phi),
\quad {\rm where} \quad
U(\phi) \equiv\sqrt{2[V(\phi) + \lambdabar ]}\;.
\end{equation}
Note that the reality of $\dot\phi$ requires $V(\phi) \geq -\lambdabar$.
Then from the scalar-field equation~\eqref{eom:phi},
the Hubble parameter becomes
\begin{equation} \label{H:V}
H = -\frac23 U'(\phi).
\end{equation}
It is not difficult to show that Eq.~\eqref{H} coincides with
Eq.~\eqref{H:V} after plugging Eq.~\eqref{dphi}. Therefore,
Eqs.~\eqref{dphi} and \eqref{H:V} provide the complete set of the
field equations with MPC.

For a given potential $U(\phi)$, in general,
one can now obtain the maximal pressure solutions (MPS)
for the scalar field and the scale factor.
Equation~\eqref{dphi} is integrated to give the scalar field,
\be\label{phieq}
\mathfrak{T}(\phi)\equiv\int^{\phi}\frac{d\varphi}{U(\varphi)}
=\int^{t}d\tau
\qquad\Longrightarrow\qquad
\phi(t)=\mathfrak{T}^{-1} (t),
\ee
and Eq.~\eqref{H:V} is integrated to give the scale factor,
\be\label{aeq}
\frac{\dot{a}}{a}=-\frac{2}{3}U'
=-\frac{2\dot{U}}{3U}
\qquad\Longrightarrow\qquad
a(t)=a_0\big[U(\phi)\big]^{-\frac23},
\ee
where $a_0$ is an integration constant.

For the massive scalar field that we consider, we obtain the MPS
from Eqs.~\eqref{phieq} and \eqref{aeq},
\be\label{MPSsq}
\phi(t) = \frac{\sqrt{2\lambdabar}}{m}\sinh[m(t-t_0)],
\qquad
a(t) = \frac{a_0}{(2\lambdabar)^{1/3}} \cosh^{-2/3}[m(t-t_0)].
\ee
The scalar field runs from $-\infty$ to $+\infty$
tracking the symmetric potential.
The Universe expands till the bouncing moment $t=t_0$, and then contracts.
The evolution is symmetric about the bouncing moment.
At the early stage ($t\ll t_0$), the Universe undergoes an exponential expansion
and no singularity is accompanied,
\be\label{MPSsqapprox}
\phi(t) \approx -\sqrt{\frac{\lambdabar}{2m^2}} \; e^{m(t-t_0)},
\qquad
a(t) \approx a_0 \left(\frac{2}{\lambdabar} \right)^{\frac{1}{3}} \; e^{\frac{2}{3}m(t-t_0)}.
\ee
At this stage, the Hubble parameter is constant and is solely determined by the mass
of the scalar field, $H=H_{\rm MPS} \approx 2m/3$.
Therefore, the curvature scale of MPS is {\it finite}
even when the energy density is very high for large $|\phi|$.
The value of $H$ is plotted in Fig.~1.
In the phase space $(\phi,\dot \phi)$,
the energy scale increases along the diagonal lines.
However, the curvature scale does not coincide with the behavior of the energy scale.

Let us discuss the stability of MPS.
We introduce the linear perturbations $\psi(t)$ and $h(t)$
for the velocities of the scalar field and the metric,
\be\label{1st}
\dot\phi =  U(\phi) \left[1+ \epsilon\psi(t)\right],
\qquad
H = -\frac23 U'(\phi) \left[1+ \epsilon h(t)\right],
\ee
and consider the field equations in the linear order in $\epsilon$.
Plugging Eq.~\eqref{1st} into the Eqs.~\eqref{eom:phi} and \eqref{H},
we get
\begin{equation}\label{dpsi}
\dot \psi -2 U' h =0,
\qquad
h = \Big(-\frac{2}{3}+\sqrt{\frac{2}{3\kappa}}\frac{1}{U'}\Big)\psi.
\end{equation}
Combining these two equations, we get
\be
\frac{\dot\psi}{\psi} = -\frac43 U' + 2\sqrt{\frac{2}{3\kappa}}.
\ee
Using the relation $UU'=\dot U$ from Eq.~\eqref{dphi},
we finally get
\begin{equation}\label{psi:1}
\psi = \psi_0 \big[U(\phi)\big]^{-\frac43} \; e^{t/t_c},
\quad
h = \psi_0 \left[-\frac{2}{3}+\sqrt{\frac{2}{3\kappa}}\frac{1}{U'(\phi)}\right]
\big[U(\phi)\big]^{-\frac43}\; e^{t/t_c},
\end{equation}
where $t_c =\sqrt{3\kappa/8}$.
For MPS of the massive scalar field in Eq.~\eqref{MPSsq},
the linear perturbations for $\dot\phi$ and $H$ in Eq.~\eqref{1st}
become
\begin{align}
U\psi &= \psi_0
 (2\lambdabar)^{-1/6} \cosh^{-1/3}[m(t-t_0)] e^{t/t_c}
 \propto
 e^{\left( \frac{m}{3} + \sqrt{\frac{8}{3\kappa}} \right)t}
 \qquad\mbox{  (as $t \to -\infty$) }, \label{pert} \\
U'h &= \psi_0
 (2\lambdabar)^{-2/3} \left\{ \sqrt{\frac{2}{3\kappa}} -\frac{2m}{3} \tanh[m(t-t_0)] \right\}
 \cosh^{-4/3}[m(t-t_0)] e^{t/t_c}
 \propto
 e^{\left( \frac{4m}{3} + \sqrt{\frac{8}{3\kappa}} \right)t}
 \qquad\mbox{  (as $t \to -\infty$) } .
\end{align}
Both of the perturbations grow in time,
so MPS is unstable.
Below, we shall show that the Universe starts from MPS at high energy,
and evolves to the usual inflationary stage due to this instability.

\begin{figure}[tbph]
\begin{center}
\begin{tabular}{cc}
\includegraphics[width=.45\linewidth,origin=tl]{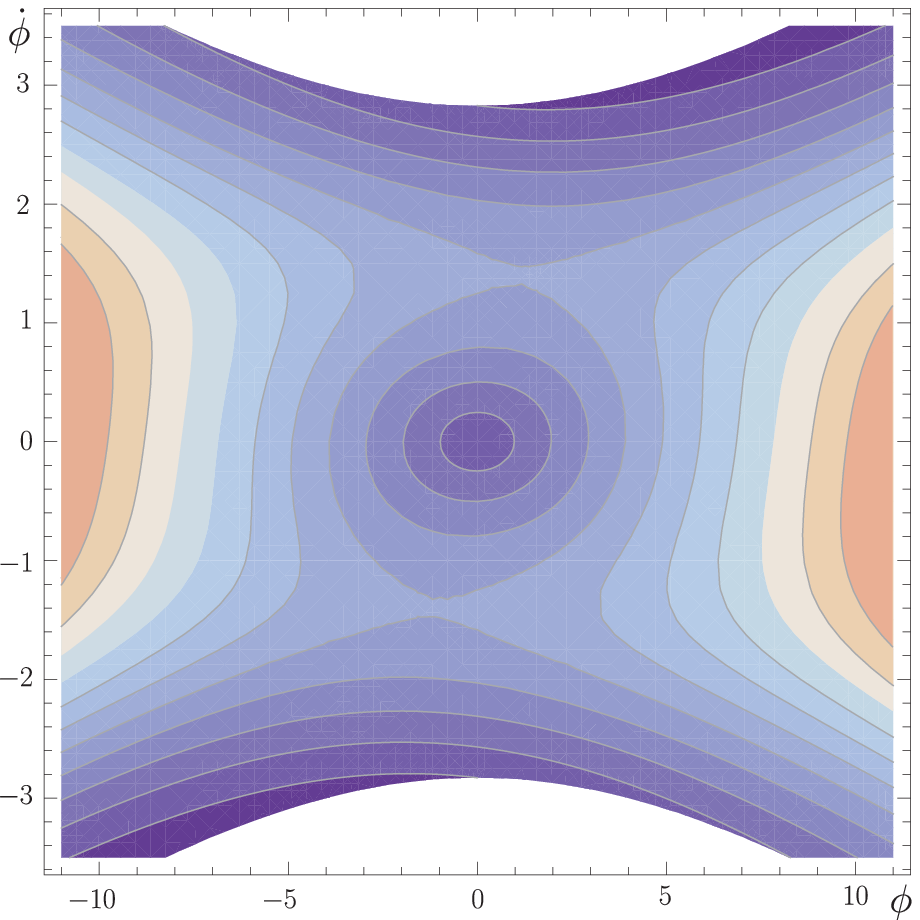}
 &\quad\includegraphics[width=.5\linewidth,origin=tl]{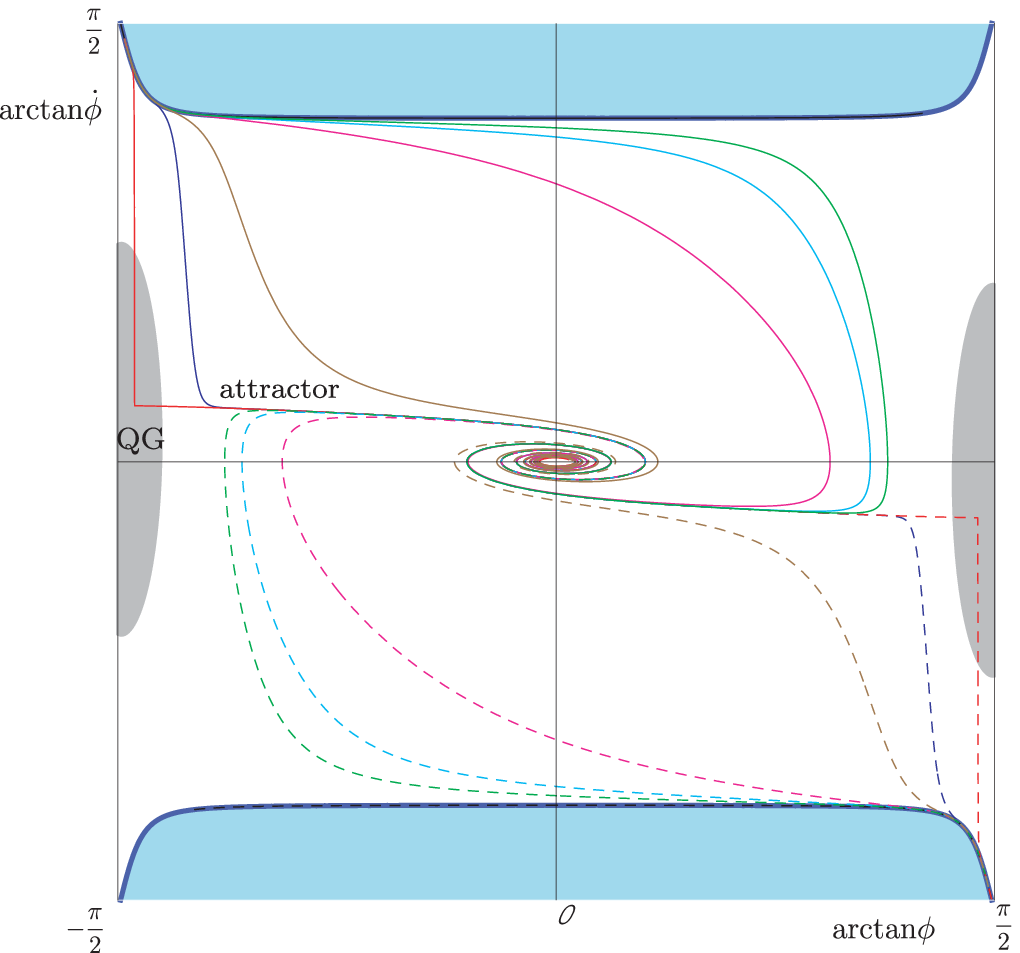}
\end{tabular}
\end{center}
\caption{Plot of the Hubble parameter $H$ (left panel) and the phase
diagram $(\phi,\dot\phi)$ (right panel) for various initial data
with $m=1/4$, $\kappa = 1/4$, and $\lambda=1$. In the left panel,
$H=0$ at the very center. The contours are separated by $\Delta H
=0.1$. The blue color represents small $H$. The reddest part denotes
the region of $H> 1$. The intermediate parts denote the region of
$0.1 \leq H \leq 1$. The region in white is physically forbidden. In
the right panel, the blue-shaded region denotes the physically
forbidden region bounded by MPS (black line). The other curves
denote the evolution of the perturbed scalar field from MPS with
other initial conditions. The solid curves denote the evolution
starting from left top, and the dashed curves denote the evolution
starting from right bottom. The attractor trajectory is the curve to
which the nearby paths converge. The Universe undergoes chaotic
inflation along the attractor trajectory until the scalar field
becomes small enough. The high-curvature regime is denoted by the
shaded region QG.}\label{fig:phi}
\end{figure}

Now let us discuss the whole picture of the evolution of the Universe (EU)
which is plotted in the phase diagram in Fig.~1 based on numerical calculations.
Consider EU of a high-energy MPS with a small perturbation.
At the early stage of EU, the Universe is closer to MPS.
As time elapses, the perturbation grows exponentially and $|\dot\phi|$ decreases.
Once EU departs from MPS considerably,
the Universe enters into the first slow-roll regime
where $|\dot\phi|$ rapidly drops.
Soon after, the Universe settles down to the second slow-roll regime which
is described by the {\it attractor} solution~\eqref{slowsol},
and undergoes the usual chaotic inflationary expansion.
Finally the Universe exits from the inflationary stage,
and the scalar field starts to oscillate about the minimum of the potential.

\vspace{10pt}
\noindent There are a few distinct features in the phase diagram.

(i) There are forbidden regions bounded by the hyperbolic MPC.

(ii) All the evolution paths of the Universe start from left top, or from right bottom in the diagram.

(iii) The high-curvature region is much suppressed than that in GR.
(Remember that the curvature scale for MPS\\\hspace*{2.7em} is
constant.)

(iv) There exists an attractor solution to which the evolution paths departed from MPS converge.

(v) All the paths but MPS settle down to the center of the diagram.
\vspace{10pt}

\noindent
The features (iv) and (v) are nothing but the familiar properties in the chaotic inflation model.
The features (i), (ii), and (iii) are new and provide a resolution to various problems of the inflation model.

In GR, the quantum theory of gravity is required
when the energy scale is larger than the Planck scale,
$3H^2 = \rho = K+V > M_P^2$, where $K=\dot\phi^2/2$ and $V=m^2\phi^2/2$.
(This quantum-gravity regime would correspond to a region outside an ellipse
if it were plotted in the phase diagram.)
For the chaotic inflation in GR, the pre-inflationary period is dominated
by the kinetic energy (K) as mentioned below Eq.~\eqref{df}.
Therefore, the pre-inflationary period inevitably enters the quantum-gravity regime.

However in the EiBI theory, the quantum-gravity regime is
considerably suppressed.
The kinetic energy is bounded
by MPS for a given value of the scalar field.
The region of $K-V> \lambdabar$ in the phase diagram is forbidden.
The curvature scale of MPS is $\sim {\cal O}(m)$,
so the quantum treatment of gravity is not necessary
in the region of $K-V \sim \lambdabar$.
While EU joins the attractor after departing from MPS,
the spacetime curvature increases to $H^2\sim m^2 \phi_0^2$.
As long as $|\phi_0|< m^{-1}$, the curvature is not comparable to the Planck scale.
In this sense, MPS provides a {\it natural} initial state for inflation
without quantum gravity.
Only in the region of $V> M_P^2$ and $V\gg K$,
the curvature scale becomes large and the quantum treatment is required
(QG in Fig.~\ref{fig:phi}).
However, this region can be avoided for the evolution paths of EU satisfying $|\phi_0|< m^{-1}$.

In the high-curvature region,
the quantum fluctuation of the scalar field is large,
$\delta \phi \sim {\cal O}(H)$.
The corresponding quantum fluctuation of the field velocity is even larger,
$\delta\dot\phi \sim \delta\phi/\delta t \sim {\cal O}(H^2)$.
This large fluctuation pushes the state in the high-curvature region
to the near-MPS region in which the classical EiBI gravity plays
rather than quantum gravity.
Therefore, the quantum fluctuation will lead the Universe
to follow the path of MPS rather than that of the quantum theory.

In Ref.~\cite{EscamillaRivera:2012vz},
the authors showed that the tensor perturbation grows linearly
in the early Universe filled with radiation
as $h_{ij}\propto A \delta \eta= A/a(t_e) \times \delta t$,
where $A$ is a constant, $t_e$ is an early moment,
and $\delta\eta=\eta-\eta_e$ is the increment of the conformal time.
For MPS, the tensor perturbation grows in a similar manner.
However, combined with the exponential growth of the MPS perturbation investigated earlier,
the problem of the tensor perturbation is naturally resolved.
Consider the evolution of tensor perturbation on EU.
For the perturbation to be well defined at $t=t_e$,
we need $A/a(t_e) = \epsilon$, where $\epsilon$ is a small parameter of order $O(m)$.
The time scale for the tensor perturbation to destroy EU
is given by $\delta t_{\rm T} \sim 1/\epsilon$.
On the other hand, following Eq.~\eqref{pert},
the MPS perturbation drives EU to the inflationary stage
in the time scale $\delta t_{\rm MPS}\sim (m/3+\sqrt{8/3\kappa})^{-1}$.
Once we have $\delta t_{\rm T} > \delta t_{\rm MPS}$,
which will be satisfied if $m \ll  1/\sqrt{\kappa}$,
the tensor perturbation fails to grow enough
simply because EU deviates from MPS exponentially fast.
Once EU settles down to the attractor solution,
the tensor mode becomes oscillatory.
Therefore, the whole picture in Fig.~\ref{fig:phi} will be kept,
and the problem of the tensor instability is resolved.

The study of the scalar perturbation will require a more rigorous
elaboration. It is very important to develop the formalism in the
EiBI theory, and will be interesting to apply it to investigating
this model. First, the growth of the scalar perturbation for MPS
needs to be slower than that of the MPS perturbation. Second, the
study of the scalar perturbation produced during the inflationary
period in this theory is very important in testing the model
compared with the observed density perturbation. We will get back to
these in near future.

In summary, we investigated the evolution of the Universe
driven by a scalar field with a quadratic potential
in Eddington-inspired Born-Infeld gravity.
When the energy density is high,
the maximal pressure condition is achieved,
for which the Universe undergoes an exponential expansion
from a nonsingular initial state.
Although the energy density is high (in fact, is not even bounded from above) during this period,
the curvature scale remains constant, $H=H_{\rm MPS} \approx 2m/3$.
This state is unstable under perturbations
and evolves to the inflationary period.
The succeeding inflation feature is the same with the ordinary chaotic inflation in GR,
but it is not quite chaotic since the pre-inflationary stage can have a finite low curvature.
Therefore, our model provides a natural precursor of inflation.
It is not clear yet whether the EiBI theory is fundamental or effective,
and whether or not this theory requires the quantum-gravity regime in the end.
However, what we can say is that quantum gravity
is not necessary in understanding the pre-inflationary stage in this model.

\section*{Acknowledgement}
This work was supported by the Korea Research Foundation (KRF) grants
funded by the Korea government (ME) NRF-2012R1A1A2006136 (I.C.) and
No. 2010-0011308 (H.K.).

\end{document}